\documentclass[pdflatex,sn-mathphys-num]{sn-jnl}

\usepackage{graphicx}%
\usepackage{multirow}%
\usepackage{amsmath,amssymb,amsfonts}%
\usepackage{amsthm}%
\usepackage{mathrsfs}%
\usepackage[title]{appendix}%
\usepackage{xcolor}%
\usepackage{textcomp}%
\usepackage{manyfoot}%
\usepackage{booktabs}%
\usepackage{algorithm}%
\usepackage{algorithmicx}%
\usepackage{algpseudocode}%
\usepackage{listings}%

\theoremstyle{thmstyleone}%
\theoremstyle{thmstyletwo}%

\theoremstyle{thmstylethree}%

\begin{document}

\title[Article Title]{Unified Approach to Portfolio Optimization using the `Gain Probability Density Function' and Applications}


\author[1]{\fnm{Jean-Patrick} \sur{Mascomère}}\email{jean-patrick.mascomere@totalenergies.com}

\author[1]{\fnm{Jérémie} \sur{Messud}}\email{jeremie.messud@totalenergies.com}

\author[1]{\fnm{Yagnik} \sur{Chatterjee}}\email{yagnik.chatterjee@totalenergies.com}

\author[1]{\fnm{Isabel} \sur{Barros Garcia}}\email{isabel.barros-garcia@totalenergies.com}

\affil[1]{\orgname{TotalEnergies}, \orgaddress{\street{Tour Coupole - 2 place Jean Millier} \city{92078 Paris La Défense cedex}, \country{France}}}


\abstract{This article proposes a unified framework for portfolio optimization (PO), recognizing an object called the `gain probability density function (PDF)' as the fundamental object of the problem from which any objective function could be derived. 
The gain PDF has the advantage of being 1-dimensional for any given portfolio and thus is easy to visualize and interpret.
The framework allows us to naturally incorporate all existing approaches (Markowitz, CVaR-deviation, higher moments...) and represents an interesting basis to develop new approaches. 
It leads us to propose a method to directly match a target PDF defined by the portfolio manager, giving them maximal control on the PO problem and moving beyond approaches that focus only on expected return and risk.
As an example, we develop an application involving a new objective function to control high profits, to be applied after a conventional PO (including expected return and risk criteria) and thus leading to sub-optimality w.r.t. the conventional objective function. 
We then propose a methodology to quantify a cost associated with this optimality deviation in a common budget unit, providing a meaningful information to portfolio managers.
Numerical experiments considering portfolios with energy-producing assets illustrate our approach. The framework is flexible and can be applied to other sectors (financial assets, etc).}





\maketitle

\tableofcontents

\section{Introduction}\label{sec1}

Modern single-period portfolio optimization (PO) started in 1952 with Markowitz’s pioneering paper \citep{markowitz1952portfolio}, which introduced the idea that investors should allocate their investments across a set of assets by balancing expected return and risk.

The PO approach is generally based on a statistical model of information related to the considered assets and the definition of a gain function that represents a portfolio 'satisfaction criterion' \citep{daniel2025portfolio,gunjan2023brief,meucci2005risk}. From these elements, statistical gain values can be computed for any given portfolio, related to a portfolio gain distribution.
Two application examples are:
\begin{itemize}
    \item Short-term financial trading, where the statistical information usually represents the return of each asset, and the gain function represents the portfolio total return, which is linear w.r.t. the portfolio asset shares \citep{chaweewanchon2022markowitz,krokhmal2002portfolio,gatfaoui2019diversifying}.
    \item Long term investment in `physical assets' (like energy-producing assets...) \citep{jano2017investment, reus2018retail, tietjen2016investment}, where the statistical information can represent the return as well as the cost related to each asset (both being uncertain in the long term), and the gain function might represent the portfolio return on investment (ROI), which is non-linear w.r.t. the portfolio asset shares.
\end{itemize}

After modeling the statistical information of the assets and defining the gain function, it is essential to establish profit and risk measures. These measures are typically expressed as expectations computed from the statistical gain values. By combining them within a multi-objective optimization, one obtains an efficient frontier (or Pareto front), which provides a set of optimal portfolios balancing expected risk and profit \citep{daniel2025portfolio}.
The portfolio manager then post-selects the most suitable portfolio according to additional criteria  \citep{xidonas2014multiobjective,bailey2012sharpe,grodzevich2006normalization}.

The most widely used profit measure is the expectation of the statistical gain values, or first-order moment of these values.
For the risk measure, various options exist:
\begin{itemize}
     \item The emblematic work of Markowitz \citep{markowitz1952portfolio} considered the second-order moment of the portfolio gain distribution, i.e. the variance of these values, in a framework where the gain function is linear. Even if Markowitz did not formulate his risk measure directly in terms of the portfilio gain distribution but using the asset information covariance matrix, both formulations are equivalent. Note, however, that the formulation used by Markowitz does not provide an obvious generalization of his risk measure to non-linear gain functions. Furthermore, the Markowitz risk measure has drawbacks. It is optimal only for a gaussian statistical model of asset information, which does not necessarily hold. Also, a risk measure that limits the profit-side variance (in addition to limiting the loss-side variance) may not represent what is desired by portfolio managers. This led Markowitz to propose the semi-variance measure of risk, that limits only the loss-side variance \citep{markovitz1959portfolio}. Still, the gaussian hypothesis is underlying.
     \item Thus, more sophisticated risk measures have been proposed, such as the value-at-risk (VaR) \citep{metrics1996jp} and the conditional-value-at-risk (CVaR) \citep{mcneil2015quantitative,rockafellar2000optimization}. The VaR emerged in the late 1980s, measuring the maximum loss within a specified confidence level and surpassing the gaussian hypothesis.
     The CVaR considers the expectation of the portfolio gain distribution beyond the VaR and has the nice feature to be sub-additive contrary to VaR \citep{artzner1999coherent}. It can be used to compute a loss-side deviation known as the CVaR-deviation \citep{rockafellar2002deviation}. 
\end{itemize}

To the best of our knowledge, the most used risk measures are all obtained from expectation values \citep{meucci2005risk}, which may not be enough to characterize the information contained in the gain distribution. Indeed, these measures typically capture only a single mode interpretation.
To remedy this, it has been proposed to consider higher-order moments of the gain distribution, e.g. skewness (third-order moment) \citep{zhai2018mean,konno1993mean} and kurtosis (fourth-order moment) \citep{lai2006mean,de2003incorporating}, that allow to better account for asymmetry and flatness in the gain distribution, but these methods remain quite limited in the general case as they only add few additional (scalar) expectation values in the objective function. 

The approach we propose in this article allows us to unify these frameworks and incorporate more information contained in the statistical gain values. It still begins with the statistical information available on the assets but also leverages the fundamental properties of the statistical framework. Namely the gain function of a given portfolio is recognized as a random variable (r.v.) whose probability density function (PDF) is the fundamental object of the PO problem from which any objective function shall be derived. We call this object the gain PDF, which has the advantage of being 1-dimensional for any portfolio. The framework allows us to naturally incorporates all existing approaches (Markowitz, CVaR-deviation, higher moments...) and represents an interesting basis to develop new approaches. It indeed leads us to propose a method to directly match a target PDF defined by the portfolio manager, giving maximal control on the PO problem.
This approach can be particularly useful when portfolio managers have specific requirements on the portfolio composition; e.g., we develop an application involving a new objective function to control high profits, to be applied after a conventional PO (including a risk or loss-oriented criterion) and thus leading to sub-optimality w.r.t. the conventional objective function. 
We then propose a methodology to quantify a cost associated with this optimality deviation in a common budget unit, providing a meaningful information to portfolio managers.

The article is structured as follows. The second section clarifies the statistical assumptions underlying the single-period PO formulation and the necessary implications that arise from
them. The gain PDF will be introduced and, as a by-product, the  Markowitz risk measure will be naturally generalized to non-linear gain functions (like ROI...). In the third section, we constraint the gain PDF estimator from first principles and very mild assumptions, and propose an original method based on the gain PDF to match a target PDF. In the fourth section, we apply the proposed method to foster high gains, as a second step after conventional PO. The method leverages projected gradient optimization and provides a ‘perturbed portfolio’. We propose a unified framework to quantify the marginal costs associated with the `perturbed portfolio', enabling portfolio managers to justify their choices quantitatively. In the last section, we present numerical illustrations of PO considering 'physical assets' and more precisely energy-producing assets.

\section{PO Framework and Gain PDF}\label{sec2}

\subsection{General Framework}
\label{sec:gen}

The problem of PO generally consists in allocating a given budget (for example, a monetary investment in short-term financial trading, an amount of energy to be produced in long term investment in energy-producing assets, etc.) among a set of $N$ assets ($A_1,..., A_N$). The objective is to determine the proportions $\vec P = (p_1, ..., p_N)^T$ of each asset to buy in order to maximize a satisfaction criterion established by the portfolio manager.
Within this framework, we have \citep{daniel2025portfolio, meucci2005risk}:
\begin{equation}
    \sum_{n=1}^N p_n =1
    \quad,\quad
    p_n\ge 0.
\label{eq:p_eq}
\end{equation}

We suppose that information about each asset \( A_n \) at the time the portfolio is constructed can be estimated (usually from past data), and takes the form of a list of $L$ values related to various features of an asset (for example the expected return, cost, etc of an asset).
These values can be arranged into a column vector \( \vec Y_n \). The vectors \( \vec Y_n \) are concatenated into a $L\times N$ matrix \( \underline Y = [\vec Y_1 \ldots \vec Y_N] \).
A number of constraints—assumed here to be linear functions of both \( \vec P \) and \(  \underline Y \), such as maximum percentage limits for each asset— can be considered and restrict the set of feasible \( \vec P \). These constraints are expressed in the form of a matrix inequality 
\begin{equation}
 \underline M \,\ \underline Y \,\ \vec P \leq \vec C,
\label{eq:constr}
\end{equation}
where \(  \underline M \) is a matrix and \(\vec C \) is a column vector. The matrix \(  \underline M \) determines which rows (i.e., which types of information) within the \( \vec Y_n \) the constraints pertain to, and the multiplication by $\vec P$ specifies which relationships between the $p_n$ must be bounded by the elements of vector \( \vec C \).

In practice, certain information—such as the expected return and sometimes the expected cost—is not known exactly at the time the portfolio is constructed. To account for possible variability, it is common to have a set of \( S \) different values for each asset feature gathered in $ \underline Y$, corresponding to possible scenarios \citep{daniel2025portfolio, meucci2005risk}.
In practical terms, \(  \underline Y \) can therefore be considered as a scenario-dependent matrix, denoted \(  \underline Y(s) \).
The first step in PO is to model relevant data \(  \underline Y(s) \) at the given time horizon. 

To optimize the portfolio selection, the portfolio manager must define a criterion
that quantifies their satisfaction with the chosen portfolio once specific values of the portfolio features \(  \underline Y(s) \) become known (only one of the scenarios being effectively realized  after the considered time step). Let us denote by \(  \underline Z \) these values \citep{daniel2025portfolio, meucci2005risk}.
This criterion usually takes the form a scalar function that we call a gain function, denoted as \( g(\vec P,  \underline Z) \):
\begin{itemize}
    \item 
Often, the indicator chosen is a measure of the portfolio total return, which can be expressed as \citep{gunjan2023brief}:
\begin{equation}
    g(\vec P, \underline Z) = \sum_{n=1}^N p_n z_n,
\end{equation}
where \( z_n \) represents the realized return for asset \( A_n \) ($\underline Z$ is then `at least' a one-row matrix), corresponding to one of the entries of the vector \( \vec Y_n \). 
\item
Alternative gain measure, such as ROI, may also be considered without altering the subsequent framework (but numerical caution must be taken as the problem becomes non-linear). The gain function then takes the form:
\begin{equation}
g(\vec P, \underline Z) = \frac{\sum_{n=1}^N p_n z_{n,1}}{\sum_{n=1}^N p_n z_{n,2}},
\label{eq:ROI}
\end{equation}
where \( z_{n,1} \) denotes the realized return for asset \( A_n \), and \( z_{n,2} \) corresponds to the realized cost for that asset ($\underline Z$ is then `at least' a two-rows matrix).
\end{itemize}

\subsection{Statistical Interpretation Choice and Gain PDF}
\label{sec:stat_interp}

Naturally, the values of \( \underline Z \) are not known at the time of portfolio construction; information about them is only available through scenario-dependent vectors \( \vec Y_n(s) \) for each scenario \( s \). Consequently, it is necessary to adopt an interpretation to utilize the collection of scenario matrices \( \underline Y(s) \) \citep{meucci2005risk,daniel2025portfolio}. A common practice in PO is to adopt a statistical framework, which, while seemingly straightforward, is essential to the subsequent arguments. The key implications of this framework are as follows:
\begin{enumerate}
    \item	The different values of each \( \vec Y_n(s) \) across scenarios \( s \) are considered independent and identically distributed (i.i.d.) realizations of the same data random vector \( \vec{\mathbf{ Y}}_n \) (where bold notation indicates random variables).
    \item	The various gain and risk measures optimized in PO must be evaluated statistically over the scenarios indexed by $s$, as result from a specific realization does not carry particular significance.
    \item	The ordering of scenarios or realizations of the data random vector \( \vec Y_n(1), \ldots, \vec Y_n(S) \) is entirely arbitrary; thus, our calculations must be invariant to any permutation of the scenario index \( s \). While representing the data as a matrix-valued function \( \underline Y(s) \) is convenient, it is not strictly necessary.
\end{enumerate}

Given this framework, the portfolio selection process requires applying the gain function \( g(\cdot, \cdot) \) to the random matrix \( \underline{\mathbf{Y}} \). As a result, \( g(P, \underline{\mathbf{Y}}) \) becomes a real-valued r.v. \citep{meucci2005risk}, which is more tractable than when using the full matrix \( \underline Y(s) \). The probability distribution of this variable—supplemented, of course, by the imposed constraints—contains all the information needed to devise an optimal portfolio selection method.

For technical reasons that are not especially restrictive, we assume that for any \( \vec P \), this distribution has a PDF with respect to the Lebesgue measure \citep{bartle2014elements}, denoted by :
\begin{equation}
\sigma(\vec P, u),    
\end{equation}
where \( u \) is the real variable associated with the gain. Due to the choice of a statistical framework, this PDF is the fundamental object in the PO problem: all relevant quantities for defining the problem can and should be constructed from it.
We call this object the gain PDF, which has the advantage of being 1-dimensional for any portfolio. 

\subsection{Objective Function Conventional Approach}
\label{sec:conv}

The selection of optimal portfolios is carried out by optimizing an objective function, which itself is a function of the PDF $\sigma(\vec P, u)$. This objective function translates an operational choice made by the portfolio manager.

The most common choice is to seek a balance between average gains and risks, which represents a multi-objective optimization whose solution is an efficient frontier \citep{daniel2025portfolio}. 
For practical purpose, the problem can equivalently be expressed as a single-objective optimization, the objective being expressed as a linear combination:
    \begin{equation}
        F_a(\sigma)= (1-a) \times\text{gain}(\sigma) - a\times\text{risk}(\sigma),
        \label{eq:orj_orig}
    \end{equation}
where $a\in [0,1]$ represents a risk aversion weight, controlling the trade-off between the gain and risk. 
We then seek the optimal portfolios,
    \begin{eqnarray}
        & \vec P_a^* = \arg\max_{\vec P} F_a(\sigma(\vec P,.))& 
        \label{eq:PO}\\
        & s.t. \quad p_n \geq 0, \quad \sum_n p_n=1, \quad \underline M \,\ \underline Y \,\ \vec P \leq \vec C&
        .\nonumber
    \end{eqnarray}
The obtained set $\{\vec P_a^*\}$ for many (ideally all) values of $a$ are related to the efficient frontier sampling.
From posterior analysis, we finally select the specific portfolio (or value of $a$) that we prefer to finalize the choice of the definitive portfolio.

The first to propose such an approach was Markowitz \citep{markowitz1952portfolio}, who considered for $\text{gain}(\sigma)$ the expected gain expressed as a mean value of the gain function $g(.,.)$, and for $\text{risk}(\sigma)$ the variance of $g(.,.)$. Formulated in terms of the gain PDF, the Markowitz approach leads to:
\begin{equation}
    \text{gain}(\sigma) = \int u\text{ }\sigma(\vec P,u)\text{ }du,
    \label{eq:Again}
\end{equation}
\begin{equation}
    \text{risk}_{Mark}(\sigma) = \int u^2\text{ }\sigma(\vec P,u)\text{ }du - \text{gain}(\sigma)^2,
    \label{eq:Mvar}
\end{equation}
where the variance has been considered in eq. (\ref{eq:Mvar}).
Even if equivalent to the formulation in eq. (\ref{eq:Mvar}), the original formulation of Markowitz considered a linear gain function $g(.,.)$ and it is not straightforward in his formulation to generalize the variance term to a non-linear gain function like a ROI, eq. (\ref{eq:ROI}).
Interestingly, the formulation in terms of the gain PDF naturally generalizes the Markowitz variance, eq. (\ref{eq:Mvar}), to non-linear gain functions.
So, the formulation in terms of the gain PDF seems more general.
Indeed, due to the choice of dealing with a statistical framework, the gain PDF represents the object containing the most exhaustive information on the PO problem. 

As mentioned in the introduction, the Markowitz risk measured as a variance leads to various pathologies and many proposal have been done to improve this measure.
Among these, the CVaR-deviation \citep{rockafellar2002deviation} allows us to compute a loss-side deviation from the average gains:
\begin{equation}
    \text{risk}_{Cdev_\beta}(\sigma) = CVaR_\beta(\sigma) - gain(\sigma),
    \label{eq:Rcvar-dev}
\end{equation}
where CVaR seeks to quantify the average risk of losses related to a confidence level $1-\beta$ with $\beta\in]0,1[$ (in practice $\beta$ is chosen close to one so that only low probability events but with strong losses are accounted for) \citep{rockafellar2000optimization,rockafellar2002deviation}:
\begin{equation}
    CVaR_\beta(\sigma) =
    \min_{\alpha}\alpha+ \frac{1}{1-\beta}\int ReLU(u-\alpha)\text{ }\sigma(\vec P,u)\text{ }du.
    \label{eq:Rcvar}
\end{equation}

These examples illustrate the central role of the gain PDF $\sigma(.,.)$ in conventional PO and the fact that any objective can be expressed as a function of this object.
However, we notice that conventional PO methods extract only a limited set of numerical values with business significance to define the objective function, through expectations, allowing to capture only limited features of the gain PDF.
As mentioned in the introduction, it has been proposed to consider few higher-order moments of the gain PDF but this still remains limited to a few additional average quantities.
Accounting for more exhaustive information contained in the gain PDF thus seems to remain an open question.

\section{New Approach Leveraging the Gain PDF}

\subsection{Objective Function}

We now go further and propose an innovative and flexible approach by granting the portfolio manager maximum freedom to express their expectations regarding the distribution of profits and losses. Specifically, this means allowing the user to define their own target gain PDF, denoted as
\begin{equation}
    \sigma_t(u),
\end{equation}
which represents the ideal PDF they would like to achieve (which is not necessarily an easy task but a constructive method will be proposed further).
In practice, the goal will be to approximate this ideal target, since the finite number of degrees of freedom (the $p_n$ together with the constraints) do not give enough flexibility to attain exactly any arbitrary PDF.

An objective function to achieve this task must be a measure of the mismatch between the gain PDF and this ideally desired PDF, called the target PDF in the following. This measure, which we seek to minimize, quantifies how close the gain PDF is to the target PDF. Clearly, this approach provides the portfolio manager with a richer means of expressing their optimality requirements compared to previous methods, which relied on only few (usually two) expectations computed using the gain PDF.

It then remains to choose a suitable discrepancy measure  $D(\sigma(\vec P, \cdot)\,;\, \sigma_t(\cdot))$ between the gain distribution produced by the portfolio and the target distribution. The PO problem can then be formulated as:
\begin{eqnarray}
    &\vec P^*_t = \arg\min_{\vec P} =D\big(\sigma(\vec P, \cdot)\,;\, \sigma_t(\cdot)\big)& 
    \label{eq:newPO}\\
        & s.t. \quad p_n \geq 0, \quad \sum_n p_n=1,  \quad\underline M \,\ \underline Y \,\ \vec P \leq \vec C.&
    \nonumber
\end{eqnarray}

The discrepancy measure must be carefully selected, as the obtained optimal portfolio $\vec P^*_t$ will depend on this choice. 
Various choices are possible, each with their own characteristics:
\begin{itemize}
    \item 
The Kullback-Leibler divergence \citep{kullback1951information} allows us to evaluate a mismatch between two PDFs:
\begin{equation}
    D\big(\sigma(\vec P,.);\sigma_t(.)\big)= \int_{-\infty}^{\infty} \sigma_t(u)\log\bigg(\frac{\sigma(\vec P,u)}{\sigma_t(u)}\bigg) du.
\end{equation}
Despite its widespread use—mainly due to its low computational cost—the Kullback-Leibler divergence is not the most suitable for our problem. Indeed, it has the drawback of not allowing a weighting of the matching in the $u$ direction. In other terms, the user cannot constrain to better match only a specific part of the PDF, but rather must constrain the entire PDF, which requires a very precise knowledge of the target (thus of the solution, which is inconsistent...) to achieve this.
    \item
Another option is to use a Wasserstein distance \citep{villani2009wasserstein}, but it is more computationally expensive.
    \item
Finally, an $L_p$ norm \citep{rachev1995probability} raised to the power $p$ (with $p \geq 1$) can also be used:
\begin{equation}
    D\big(\sigma(\vec P,.);\sigma_t(.)\big) = \int_{-\infty}^{\infty} \theta(u)\text{ }|\sigma(\vec P,u)- \sigma_t(u)|^p \text{ }du.
    \label{eq:L2}
\end{equation}
It has the advantage of being inexpensive to compute and, more importantly, allows us to naturally introduce a weighting function $\theta(u)$ to foster the matching of only specific parts of the target PDF in the optimization. 
\end{itemize}
We choose the latter discrepancy measure for $D(\sigma(\vec P,.);\sigma_t(.))$ with $p=2$ (squared L2-norm) \citep{stein2009real} and $\theta(u)\in[0,1]$ (e.g. the sigmoid function) to specify which part of the PDF is important to approximate as closely as possible ($\theta(u)$ close to $1$), and which part of the PDF can be given more leeway ($\theta(u)$ close to $0$).
In other terms, the target PDF must be precisely defined for the part where $\theta(u)$ close to $1$ only. An application targeting high-gains (which are not accounted for in the conventional PO scheme) will be proposed further.

\subsection{Estimator Choice}

The formulations above still do not allow us to solve the PO problem, i.e. to find the optimal portfolios through eq. (\ref{eq:PO}) or eq. (\ref{eq:newPO}).
Indeed, the gain PDF $\sigma(\vec P, u)$ form is not known a priori. 
It is necessary to specify how we will express the gain PDF as a function of the data available to us, that is, to choose its estimation based on portfolio assets features values from each scenario, which requires an explicit statistical model. 

To constrain the form of an estimator of the gain PDF, we consider first that the order in which the scenarios appear is arbitrary, as explained in section \ref{sec:stat_interp}, so that the estimator must be a function of $g(\vec P, \underline Y(s))$ that is invariant under permutation of the scenarios indexed by $s$. The distribution:
\begin{equation}
h(\vec P, u) = \frac{1}{S} \sum_{s=1}^S \delta\Big(u - g(\vec P, \underline Y(s))\Big),
\label{eq:h}
\end{equation}
therefore represents a sufficient statistics \citep{lehmann1998theory}. We call it the `gain histogram' by slight abuse of language.
The convolution of $h$ with a rectangular function of defined width gives the gain histogram in the common sense of the term, our gain histogram representing the limit of the common gain histogram when the rectangle's width is infinitesimal \citep{silverman2018density}.
All our estimators must be functions of the gain histogram $h(\vec P, u)$.

Secondly, it is important to differentiate two cases:
\begin{itemize}
    \item 
When it comes to estimating expectation values, it is natural, for large number of available scenarios $S$, to use the empirical means or averages over the scenarios, related to considering eq. (\ref{eq:h}) as the probability distribution.
Let us take the example of the Markowitz \citep{markovitz1959portfolio} or CVaR-based \citep{rockafellar2000optimization} components of the objective function that are expressed as expectations, remind eqs. (\ref{eq:Again}), (\ref{eq:Mvar}) and (\ref{eq:Rcvar}).
Using eq. (\ref{eq:h}) allows us to recover the conventional formulations in terms of averages over scenarios \citep{markowitz1952portfolio, rockafellar2000optimization,rockafellar2002deviation}:
\begin{equation}
    \text{gain}(\sigma(\vec P,.)) \approx\text{gain}(h(\vec P,.)) = \frac{1}{S} \sum_{s=1}^S g(\vec P, \underline Y(s)),
    \label{eq:Again2}
\end{equation}
\begin{equation}
    \text{risk}_{Mark}(\sigma(\vec P,.)) \approx \text{risk}_{Mark}(h(\vec P,.)) = {\frac{1}{S} \sum_{s=1}^S g(\vec P, \underline Y(s))^2 - \text{gain}(\sigma)^2},
    \label{eq:Mvar2}
\end{equation}
\begin{equation}
    CVaR_\beta(\sigma(\vec P,.)) \approx CVaR_\beta(h(\vec P,.)) =
    \min_{\alpha}\alpha+ \frac{1}{1-\beta}\frac{1}{S} \sum_{s=1}^S ReLU(g(\vec P, \underline Y(s))-\alpha).
    \label{eq:Rcvar2}
\end{equation}
Again, eq. (\ref{eq:Mvar2}) naturally generalizes the Markowitz scheme (including the expectations estimations) to non-linear gain functions like ROI-based ones.
\item
When it comes to estimating an entire function, here the gain PDF, things are not that simple ($h(\vec P,.)$ cannot be directly taken as the estimator) and several possibilities exist.
We now demonstrate that we can drastically reduce the choice of possible estimators using arguments that stem from the problem itself and from the hypotheses we have already made.
\end{itemize}

The estimation of the gain PDF is represented by an operator acting on the gain histogram that we denote as $G(h)(v)$.
To be an estimator of a PDF, $G(h)(v)$ must be positive and its integral must be equal to 1. We will therefore seek $G$ in the form:
\begin{equation}
G(h) = \frac{F(h)}{\int F(h)(v) dv}.
\label{eq:G}
\end{equation}

Under very reasonable hypotheses explicated in the demonstration below, we show that $F$ must be expressed as a convolution product
\footnote{
Translation invariance suggests that, instead of the gain histogram, we could consider its Fourier Transform as the fundamental statistic of the problem and thus also the gain characteristic function instead of the gain PDF. Detailing this goes beyond the scope of the article.
}:
\begin{equation}
    F(h)(v) = \int K(v-u)\text{ }h(u)\text{ }du,
\end{equation}
where $K(.)$ represents a chosen the convolution kernel \citep{hardle2012smoothing}.
For instance and as already mentioned, if $K(.)$ is chosen as a rectangular function, $F(h)(v)$ a gain histogram in the common sense of the term.

$G(h)$ can therefore be considered as an estimator of the gain PDF using the kernel method \citep{silverman2018density}, which in itself is not a novel technique. However, it is worth noting that the form of the gain PDF estimator is quite constrained by the problem's structure, as well as some fairly natural additional assumptions.
The kernel $K(.)$ is thus the only degree of freedom of the gain PDF estimator.

Note that we could further constrain $K(.)$ and require, for reasons of symmetry, that it must be sought within the class of even functions with quasi-bounded support. Finally, kernels such as Gaussian, Triangle, 
etc \citep{hardle2012smoothing}. The final result should not depend much on the choice of the kernel to first order (which we verified in many practical cases).\\

\textit{$\Box$ Demonstration that $K(.)$ must be a convolution kernel under mild hypotheses.} We can constrain the $G(h)$ based on two observations that seem naturally suited to our problem:
\begin{itemize}
    \item Adding the same value \( w \) to each samples \( u(s) \) of the gain PDF $\sigma(.,u)$ should corresponds to a samples drawn from a PDF \( \sigma'(.,u) = \sigma(.,u-w) \). The related gain histogram is \( h_w(.,u) = h(.,u-w) \), which represents a translated version by an amount \( w \). \( G \) must thus estimate \( \sigma_w \) as effectively as \( \sigma \) and it is legitimate to require that \( G \) (and thus \( F \)) must be a translation-invariant operator:
    \begin{equation}
        G\Big(h(\cdot, \star- w)\Big)(v) = G(h(\cdot,\star))(v - w)
    \end{equation}
\item The scenarios or samples should be interpretable as drawn from a mixture of PDFs. More precisely, we suppose that if the scenarios consist of two types of samples drawn from the PDFs \( \sigma_1 \) and \( \sigma_2 \) in proportions \( \gamma \) and \( (1 - \gamma) \), the overall samples can be considered as drawn from the single PDF \( \sigma_3 = \gamma  \sigma_1 + (1-\gamma) \sigma_2 \). This leads to \( h_3 = \gamma h_1 + (1 - \gamma) h_2 \) for the corresponding  gain histogram. 
It is legitimate to require that:
\begin{equation}
G(h_3) = G\Big(\gamma h_1 + (1-\gamma) h_2\Big) = \gamma G(h_1) + (1-\gamma) G(h_2),
\label{eq:estG}
\end{equation}
so that \( G(h) \) can estimate \( \sigma_3 \) using \( h_3= \gamma h_1 + (1 - \gamma) h_2 \) in addition to estimating \( \sigma_1 \) and \( \sigma_2 \) using \( h_1 \) and \( h_2 \).
The condition in eq. (\ref{eq:estG}) is satisfied if we require \( F(h) \) to be a linear operator
\footnote{
Indeed, the translation invariance combined with the linearity of \( F \) implies that the normalization coefficient in eq. (\ref{eq:G}) takes the same value for $G(\gamma h_1 + (1-\gamma) h_2)$ and $G(h)$.
}.
\end{itemize}
%
%
$F$ must therefore be sought within the class of linear operators that are equivariant under translation, so $F$ can be expressed as a convolution product. $\Box$

\subsection{Full Specification of the New Optimization Problem}
\label{sec:meth}

We now have everything needed to fully establish the objective function in eq. (\ref{eq:newPO}) related to our new proposal, from the problem data or scenarios. The steps are the following:
\begin{itemize}
    \item A target PDF $\sigma_t(v)$ is first specified by the user, that needs to be accurate only on the part the user would like to precisely match (this can be challenging for the user but a constructive method will be proposed further). 
    \item For any portfolio $\vec P$ and a chosen convolution kernel $K(.)$, an estimation $\hat\sigma(\vec P,v)$ of the gain PDF based on the form we derived for $G(h)(v)$ and the scenarios $\underline Y(s)$ is computed:
    \begin{eqnarray}\label{eq:estim}
        \hat\sigma(\vec P,v)&= & \frac{\int K(v-u)\text{ }h(\vec P,u)\text{ }du}{\alpha(\vec P)}\\
        h(\vec P, u) &=& \frac{1}{S} \sum_{s=1}^S \delta\Big(u - g(\vec P, \underline Y(s))\Big)\nonumber\\
        \alpha(\vec P)&=&\int K(v-u)\text{ }h(\vec P,u)\text{ }du\text{ }dv\nonumber
        .
    \end{eqnarray}
    \item A L2-based measure of the discrepancy between the estimator of the gain PDF and the target PDF is computed:
    \begin{equation}
        D(\hat\sigma(\vec P,.); \sigma_t(.)) = \int_{-\infty}^{\infty} \theta(v)\text{ }|\hat\sigma(\vec P,v)- \sigma_t(v)|^2 \text{ }dv.
        \label{eq:L2bis}
    \end{equation}
    This measure must be minimized under constraints to determine the optimal $\vec P_t^*$, eq. (\ref{eq:newPO}).
    $\theta(v)$ represents a user-defined sigmoid-based weight that indicates which part of the target PDF must be precisely matched.
\end{itemize}
The remaining non-trivial task is to build a pertinent target PDF, corresponding to an operational need.
We now propose a methodology and an application targeting high-gains, which are not accounted for in the conventional PO scheme.

\section{A Proposed Application: Constraining High Gains}
\label{sec:app}

\subsection{Operational Context and Options}

We wish to address the following operational issue: In conventional PO methods such as Markowitz \citep{markowitz1952portfolio, markovitz1959portfolio} or CVaR-based \citep{rockafellar2000optimization, rockafellar2002deviation} PO, the choice of a portfolio is determined by a trade-off between average return and risk. However, the portfolio manager may want to include a small portion of risky assets with high potential profits.
These assets are most probably not selected by conventional PO, as they usually carry a higher risk of significant losses and no term fostering specifically the high returns is usually included in the optimization.

Two approaches are available for redefining the objective function of the problem:
\begin{itemize}
    \item 
The first approach builds on the method proposed in section \ref{sec:conv} and eq. (\ref{eq:PO}). 
If the portfolio manager is not fully satisfied with the assets selected form an initial conventional PO step,  
a third average term in the objective function (\ref{eq:orj_orig}) might be added to integrate, for example, a CVaR-type term for high returns (in the same spirit that eq. (\ref{eq:Rcvar2}) but focused on the profit-side tail of the PDF). Similar to the risk aversion weight for the risk term, an additional weight that favors strong and risky profits must be defined.
The PO is then solved again considering this `augmented' objective function; it is to be noted that the optimization solver might be adapted and the computational cost of the optimization might be strongly increased.
Indeed, introducing an additional CVaR term focused on the return-side tail of the PDF might result in a minimax problem.
    \item
The second approach, which we propose, builds on the new method described in section \ref{sec:meth}.
We still suppose that an initial conventional PO step has been performed, and would like to explore `perturbations' around the  selected portfolio $P^*_a$. In this case, the user can compute the estimate $\hat\sigma(\vec P^*_a,v)$ of the gain PDF for the optimal portfolio, and is thus well positioned to define from it a target PDF $\sigma_t(v)$ by applying a perturbation to it, for instance increasing the profit-side of $\hat\sigma(\vec P^*_a,v)$ in a fully controlled way. Relaxing the loss-side is made possible by adapting the  sigmoid-based weight $\theta(v)$ (here close to $1$ for gains and close to $0$ for losses). Once the target $\sigma_t(v)$ defined, the problem is then posed as in section \ref{sec:meth} and eq. (\ref{eq:newPO}), with the additional aspect that we seek a corrective adjustment to the initial portfolio.
This makes an iterative local-optimization method initialized with $\vec P^*_a$ appropriate,
as described in next section,
which has the advantage of a reasonable numerical cost.
\end{itemize}

\subsection{Projected Gradient Algorithm}

Since the goal is to apply a correction to an already optimized portfolio, we consider an iterative 
projected gradient scheme \citep{bertsekas1999nonlinear}:
\begin{equation}
\vec P_{k+1} = \vec P_k -\kappa \times d\vec P_k
\quad,
\quad \vec P_{k=0}=\vec P_a^*,
\end{equation}
where $d\vec P_k$ represents a projected gradient and $\kappa$ the descent step. It is not possible to take directly for $d\vec P_k$ the gradient \citep{bertsekas1999nonlinear}:
\begin{equation}
\vec{\nabla}_{(\vec P)} D(\hat\sigma(\vec P, \cdot) ; \sigma_t(\cdot)).
\end{equation}
Indeed, the constraints in eq. (\ref{eq:newPO}) are not explicitly accounted for by a standard gradient descent, meaning that modifying $\vec P_k$ into $\vec P_{k+1}$ could (and in practice will) result in a violation of these constraints.
A solution is to use the projected gradient technique,
\begin{equation}
d\vec P_k = \underline{\text{Proj}}_k  \vec{\nabla}_{(\vec P)} D(\hat\sigma(\vec P, \cdot); \sigma_t(\cdot)),
\end{equation}
where the projector $\underline{\text{Proj}}_k$ ensures each obtained $\vec P_{k+1}$ satisfy all constraints.
The projector is defined adaptively:
\begin{itemize}
    \item First, we identify the constraints that become violated when moving from $k-1$ to $k$ in a conventional gradient scheme and denote by $d_k$ the corresponding number of constraints. These constraints must be re-identified at each iteration, leading to a non-linearity and justifying the term `adaptive'.
    \item Since all constraints we consider are linear in $\vec P$, they can be interpreted as scalar products of vectors with $\vec P$. We gather these vectors for the $p_n \geq 0$ and $\underline M \,\ \underline Y \,\ \vec P \leq \vec C$ violated constraints in a matrix $(\vec V_1, \ldots, \vec V_{c})$ (to avoid overloading the notation the dependence of the $\vec V_i$ on $k$ is not made explicit). 
    We must also respect the portfolio constraint $\sum_{n=1}^N p_n = 1$, which is interpreted as the scalar product of $\vec P$ with the vector $\vec V_0 = (1, \ldots, 1)$; the latter constraint will most probably always be violated so that $\vec V_0$ will always be part of the set.
    The set of vectors $(\vec V_0, \vec V_1, \ldots, \vec V_{d})$ generates a vector subspace $E$ spanned by $(\vec e_0, \ldots, \vec e_{d'})$ ($d'\le d$ denotes the number of independent constraints). 
    The idea is then to project the standard gradient vector onto the vector subspace orthogonal to $E_k$ (re-injecting the $k$ dependence which illustrates that the violated constraints may not always be the same from one iteration to the other), with the corresponding projector defined by:
    \begin{equation}
        \underline{\text{Proj}}_k = \underline I - \underline{Q(k)} \,\ \underline{Q(k)}^T
        \quad,\quad
        \underline{Q(k)} = (\vec e_0(k), \ldots, \vec e_{d_k'}(k))
    \end{equation}

    Then, the scalar products of the projected gradient with each of the $\vec V_i$ is ensured to be zero, so that the obtained $\vec P_{k+1}$ will not anymore violate any constraint.  
\end{itemize}
The final adaptively projected gradient scheme is:
\begin{equation}
\vec P_{k+1} = \vec P_k - \kappa \times \big(\underline I - \underline{Q(k)}\,\ \underline{Q(k)}^T\big)  \vec{\nabla}_{(\vec P_k)} D(\hat\sigma(\vec P_k, \cdot); \sigma_t(\cdot)),
\end{equation}
where the standard gradient vector is numerically estimated by finite differences.
%
The computationally more expensive point may be that the $(\vec e_0(k), \ldots, \vec e_{d_k'}(k))$ and the projector need to be recalculated at each iteration $k$ in the worst case. However, since the step size $\kappa$ is assumed to be small, it is possible that the violated constraints remain the same as the iterations proceed, and that one only needs to complete the basis of constraint vectors $\vec V_i$ defined at iteration $k$ from the violated constraints with additional vectors from newly violated constraints (and not to remove any vector $\vec e_i$). In this case, the Gram-Schmidt process \citep{strang2012linear} (which is iterative and does not require matrix inversion) allows one to retain most of the calculations from the previous step. 
For each $k$, one can indeed derive from the vectors $\vec V_c$ an orthonormal basis of $E_k$ using the Gram-Schmidt process \citep{strang2012linear} initialized with $\vec e_0\propto \vec V_0$.

We highlight that violations of the positivity constraints for $p_n$ can be handled more simply and equivalently, to reduce the computational cost. Indeed, the projector onto the subspace of these particular constraints is diagonal and thus commutes with any other. It is therefore sufficient to project the vectors $\vec V_i$ related to the other violated constraints onto this subspace (which is easily done by setting the corresponding entries to zero) before undertaking the Gram-Schmidt process \citep{strang2012linear} on these modified vectors only. The dimension of the projector, and thus the amount of computation required, is accordingly reduced.


\subsection{Evaluation of Marginal Cost}
\label{sec:marg_cost}

With the previously described adjustment, we have modified a portfolio $\vec P_a^*$ that was initially computed to satisfy a conventional PO optimality criterion, i.e. maximizing an `original' objective function as in eqs. (\ref{eq:orj_orig})-(\ref{eq:PO}).
The new portfolio $\vec P_t^*$ is obtained by considering another objective function ,
eq. (\ref{eq:newPO}), and thus must be sub-optimal according to the original objective function:
    \begin{eqnarray}\label{eq:sub_opt}
        F_a(\vec P_t^*) < F_a(\vec P_a^*),
    \end{eqnarray}
where we denote $F_a(h(\vec P, \cdot))$ by $F_a(\vec P)$ to lighten the notations.
The gap $F_a(\vec P_a^*)-F_a(\vec P_t^*)$ provides an indication of the trade-offs that are made by adopting the portfolio $\vec P_t^*$ instead of $\vec P_a^*$, as the objective function represents a business satisfaction measure, but this gap is not easy to interpret in terms of 'additional cost'. 
We propose a method that addresses this point and the following operational question: 
\textit{Can we quantify, in `additional cost' terms relevant to the portfolio manager, this loss of optimality?}

This question can be encapsulated into a general framework, where the cost is measured in a single 'budget' unit so that different portfolio modifications can be easily compared (for instance a monetary unit in financial assets cases, an energy unit in energy assets cases, etc).
To begin with, let us consider eqs. (\ref{eq:p_eq})-(\ref{eq:constr}) and make explicit the case where our portfolio problem is to allocates a given budget $B$ (or investment) among assets.
We assume that in the $\vec Y_n$ data features vectors, one of the entry quantifies the cost $y_n$ to buy one unit of asset $A_n$. 
The fundamental portfolio problem is to find the quantities $x_n$ of each asset to buy w.r.t. a given budget $B$:
\begin{equation}
    \sum_{n=1}^N x_n y_n =B
    \quad,\quad
    x_n\ge 0,
\label{eq:x_eq}
\end{equation}
instead of eq. (\ref{eq:p_eq}), with inequality constraints acting on $X$ instead of eq. (\ref{eq:constr}). Then, doing the variable change
\begin{equation}
p_n=x_n \frac{y_n}{B},
\end{equation}
leads to the notations of section \ref{sec:gen}.
This allows us to understand that the  optimal objective function values, and portfolio $\vec P^*_a$ and $\vec P^*_t$ must be considered as functions of the budget $B$:
\begin{equation}
\vec P^*_a(B)
\quad,\quad
\vec P_t^*(B).
\end{equation}
In many PO cases, the role of $B$ is crucial. This is especially true with energy assets PO where many inequality constraints are necessary (related to a a maximum possible energy production per asset) and can often be saturated, and a non-linear gain function is used (e.g. ROI). All this leads to a non-trivial dependency of $\vec P^*_a(B)$ and the objective function on $B$. 

Now that the meaning of the budget $B$ has been clarified, we come back to the original PO problem objective function,
\begin{equation}
     F_a(\vec P) = (1-a) \times \text{gain}(\vec P) + a \times \text{risk}(\vec P),
\end{equation}
where the risk term can for instance be the CVaR-deviation, $\text{risk}_{Cdev_\beta}(\vec P)$, eqs. (\ref{eq:Rcvar-dev})-(\ref{eq:Rcvar}),
and to eq. (\ref{eq:sub_opt}),
    \begin{eqnarray}
        F_a(\vec P_t^*(B)) < F_a(\vec P_a^*(B)).
    \end{eqnarray}
    
We note that the $F_a(\vec P_t^*)$ lower value could have been achieved by optimizing the initial objective function while accounting for a budget $B-\Delta B_t$:
\begin{equation}\label{eq:delta_B}
F_a\big(\vec P_t^*(B)\big) = F_a\big(\vec P_a^*(B-\Delta B_t)\big).
\end{equation}
We can then search for the $\Delta B_t$ that solves the previous equation.
Practically, we calculate the optimal portfolios \( F_a(\vec P_a^*(B-\Delta B)) \) for various $\Delta B$ values by running various optimizations considering the original objective function with a constant risk-aversion value of \( a \) (parallel computing \citep{severance2010high} is an ideal tool here). We then select the specific $\Delta B_t$ value that allows us to satisfies eq. (\ref{eq:delta_B}).

$\Delta B_t$ can be interpreted as a cost of sub-optimality, here called marginal cost, representing the common budget unit we were searching for.
Intuitively, considering an identical objective function value as in eq. (\ref{eq:delta_B}) means considering an identical business satisfaction measure.
$\Delta B_t$ can thus be interpreted as the investment amount the portfolio manager would save in the original problem to obtain a business satisfaction value identical to the one related to $\vec P_t^*(B)$.
Conversely, $\Delta B_t$ approximately gives the investment amount the portfolio manager would have to in add in the target gain PDF matching problem to obtain a business satisfaction value identical to the one related to $\vec P_a^*(B)$,
\begin{equation}
F_a\big(\vec P_t^*(B+\Delta B_t)\big) \approx F_a\big(\vec P_a^*(B)\big),
\end{equation}
which is valid in the case where $\Delta B_t$ is sufficiently small so that the linearity hypothesis holds.


This idea of assigning a marginal cost value to the modification of an optimal portfolio is all the more interesting as it can be generalized to various optimization problem parameters or features.
As an example, we can also assign a marginal cost to a modification $a'$ of any selected risk aversion parameter \( a \).
We have:
\begin{eqnarray}
    F_a(\vec P_{a'}^*(B)) < F_a(\vec P_a^*(B)),
\end{eqnarray}
and search for the marginal cost $\Delta B_{a'}$ such as
\begin{eqnarray}\label{eq:Pa'}
    F_a\big(\vec P_{a'}^*(B+\Delta B_{a'})\big) = F_a\big(\vec P_a^*(B)\big),
\end{eqnarray}
equivalent for a sufficiently small $\Delta B_{a'}$ to
\begin{eqnarray}
    F_a\big(\vec P_{a'}^*(B)\big) \approx F_a\big(\vec P_a^*(B-\Delta B_{a'})\big).
\end{eqnarray}
Computing $\Delta B_{a'}$ provides a marginal cost to the modification of an optimal portfolio related to a risk-aversion value change $a'=a + \Delta a'$. In other words, it provides a marginal cost to a risk level modification \citep{bauer2016marginal} and 
answers the following operational question:  
\textit{If we had taken a little more risk, could we have saved a significant amount of investment?}
Interestingly, the marginal cost quantifying the risk aversion parameter variation $\Delta a'$ is given in the same budget unit than in the case of the target gain PDF matching.
The method proposed here thus represents a way to quantitatively and homogeneously compare heterogeneous variations of parameters and features of the PO problem, in budget variation terms interpretable by the portfolio manager.

In practical terms, computing $\Delta B_{a'}$ can be done using the method involving parallel computing mentioned above. However, a more efficient numerical scheme can be derived considering that $\Delta B_{a'}$ is small.
A constant business satisfaction or objective function value w.r.t. a differential variation of the pair \( (B, a) \) is expressed as:
\begin{equation}
0 = dF_a(\vec P_a^*(B)) = \frac{\partial F_a(\vec P_a^*(B))}{\partial B} \, dB + \frac{\partial F_a(\vec P_a^*(B))}{\partial a} \, da
\end{equation}
Around the point \( (B, a) \), variations $dB$ of $B$ and $da$ of $a$ that maintain constant the value of the objective function are coupled: $dB = -\frac{\partial F_a(\vec P_a^*(B))}{\partial a}/\frac{\partial F_a(\vec P_a^*(B))}{\partial B}$, which leads to
\begin{equation}\label{eq:Pa'2}
\Delta B_{a'} = -\frac{\frac{\partial F_a(\vec P_a^*(B))}{\partial a}}{\frac{\partial F_a(\vec P_a^*(B))}{\partial B}} \, \Delta a'.
\end{equation}
The ratio in the previous equation is in practice computed using a finite difference method and represents a solution to eq. (\ref{eq:Pa'}). Again, eq. (\ref{eq:Pa'2}) can be used only for sufficiently small $\Delta a'$ and $\Delta B_{a'}$. For larger variations, one has to consider directly eq. (\ref{eq:Pa'}) and parallel computation.
%
\begin{figure}[h]
    \centering
    \includegraphics[width=0.61\linewidth]{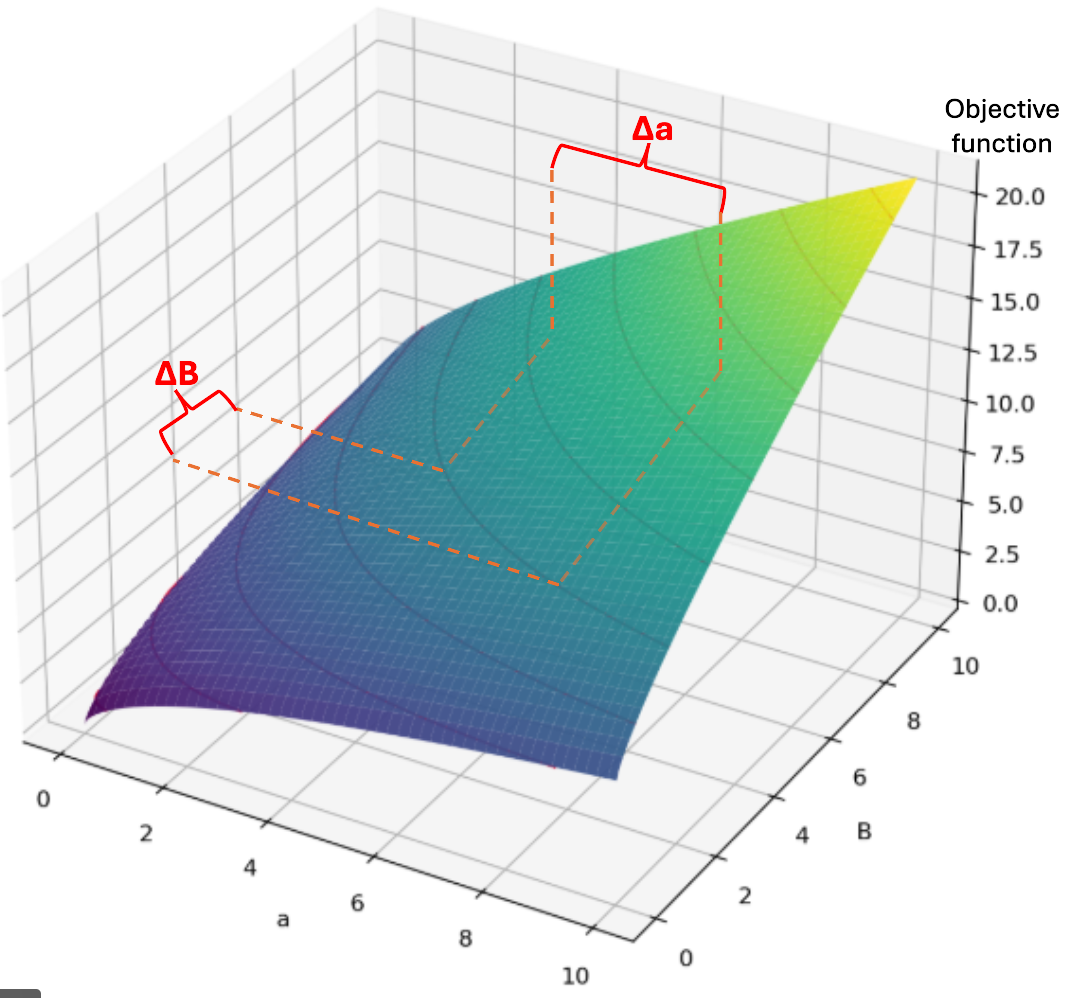}
    \caption{Objective function value landscape $F_a(\vec P_a^*(B))$ for different values of $a$ and $B$.}
    \label{fig:lands}
\end{figure}

We now go a step further and would like to compute all pairs $(B-\Delta B_{a'}, a+\Delta a')$ that satisfy eq. (\ref{eq:Pa'}), i.e. leading to a similar initial objective function or business satisfaction value than the pair $(B, a)$, requiring several independent optimizations and thus strongly benefiting from parallel computing. 
We obtain an objective function value landscape
from which iso-objective function lines related to any original $(B, a)$ value choices can be `drawn', and thus from which all $(B-\Delta B_{a'},a+\Delta a')$ values satisfying eq.  (\ref{eq:Pa'}) can be recovered.
An illustrative example is given in Fig. \ref{fig:lands}.

This method naturally extends to any kind of portfolio optimality deviations,
e.g. to evaluate the marginal cost of modifying the $C$ parameters in the inequality constraints, eq. (\ref{eq:constr}).
Again, the obtained cost will be in the same budget units than for the other optimality deviations described above, allowing to homogeneously compare heterogeneous variations of parameters and features of the PO problem, and making the result easily interpretable by the portfolio manager.

\section{Numerical Illustration}

\subsection{Energy Assets PO}

Having established the theoretical framework and proposed a method for PO leveraging the gain PDF together with marginal cost evaluation, we now illustrate how these approaches perform in practice. The following section presents numerical examples in the context of energy asset portfolios. 

Portfolio optimization in energy markets is a particularly relevant application area, given the sector’s intrinsic volatility, regulatory complexity, and technological heterogeneity. This problem is frequently explored in two main contexts: short-term trading, which focuses on navigating high-frequency price fluctuations in electricity, petroleum, and gas markets  \citep{faia2017ad, narajewski2022optimal, gatfaoui2019diversifying}, and long-term asset investment, which emphasizes diversified portfolios that integrate conventional and renewable sources while accounting for regulatory incentives and demand uncertainty  \citep{jano2017investment, reus2018retail, tietjen2016investment}. It is on the latter case that our illustrations focus.

Across these contexts, researchers have applied portfolio optimization techniques to a variety of energy markets, including petroleum products \citep{lim2020analysis}, hydro resources \citep{lemos2012hydro}, natural gas \citep{rebennack2010energy}, coal \citep{selccuklu2023electricity}, and hybrid renewable portfolios such as solar, wind, PV, and biomass \citep{maier2016risk, stempien2017addressing, tolis2011impact}. Risk is typically modeled using metrics such as VaR \citep{berleant2005electric, narajewski2022optimal}, CVaR \citep{ma2021optimal, bazmohammadi2018portfolio}, and the mean-variance approach \citep{tang2017selection, reus2018retail}. 
These approaches rely on multi-objective optimization of expectations of few functions of the gain distribution \citep{meucci2005risk}, which may not be enough to characterize the information contained in the gain PDF.
In the following numerical illustrations, we demonstrate how our unified approach exploiting directly the gain PDF addresses these limitations.

\subsection{Constraining High Gains and Evaluating Marginal Cost}

We consider synthetic data associated to energy producing assets $A_n$, e.g. electricity generation plants distributed across multiple technologies (onshore and offshore wind farms, solar plants, gas plants...) as well as various countries. Among these assets, some are classified as ‘Secure’, characterized by a scenario distributions with small variance around the mean, while others are designated as ‘Merchant’ assets, characterized by greater uncertainty and broader scenario distributions.

\begin{figure}[h]
    \centering
    \begin{minipage}{0.95\linewidth}
        \centering
        \includegraphics[width=\linewidth]{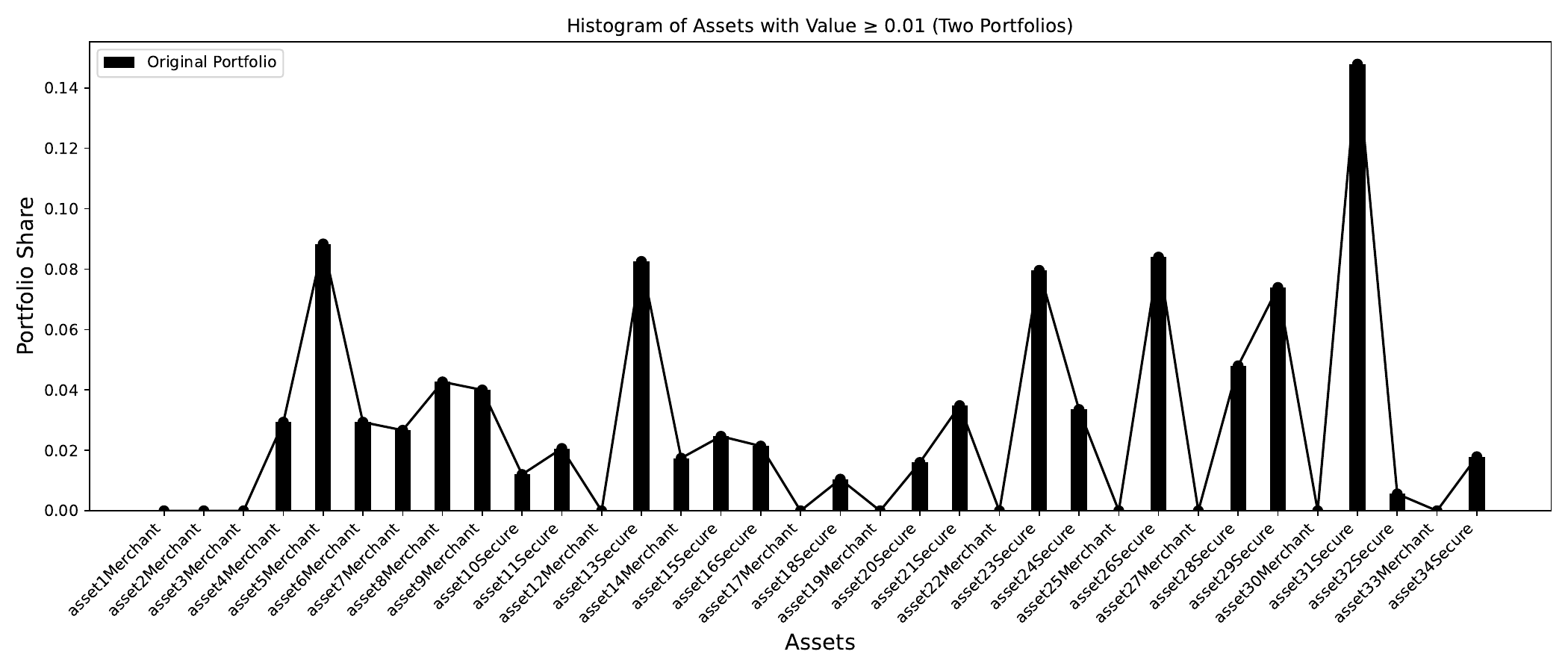}
        \caption{Original portfolio (after classical PO).}
        \label{fig:losspdf1}
    \end{minipage}
    \hfill
    \begin{minipage}{0.8\linewidth}
        \centering
        \includegraphics[width=\linewidth]{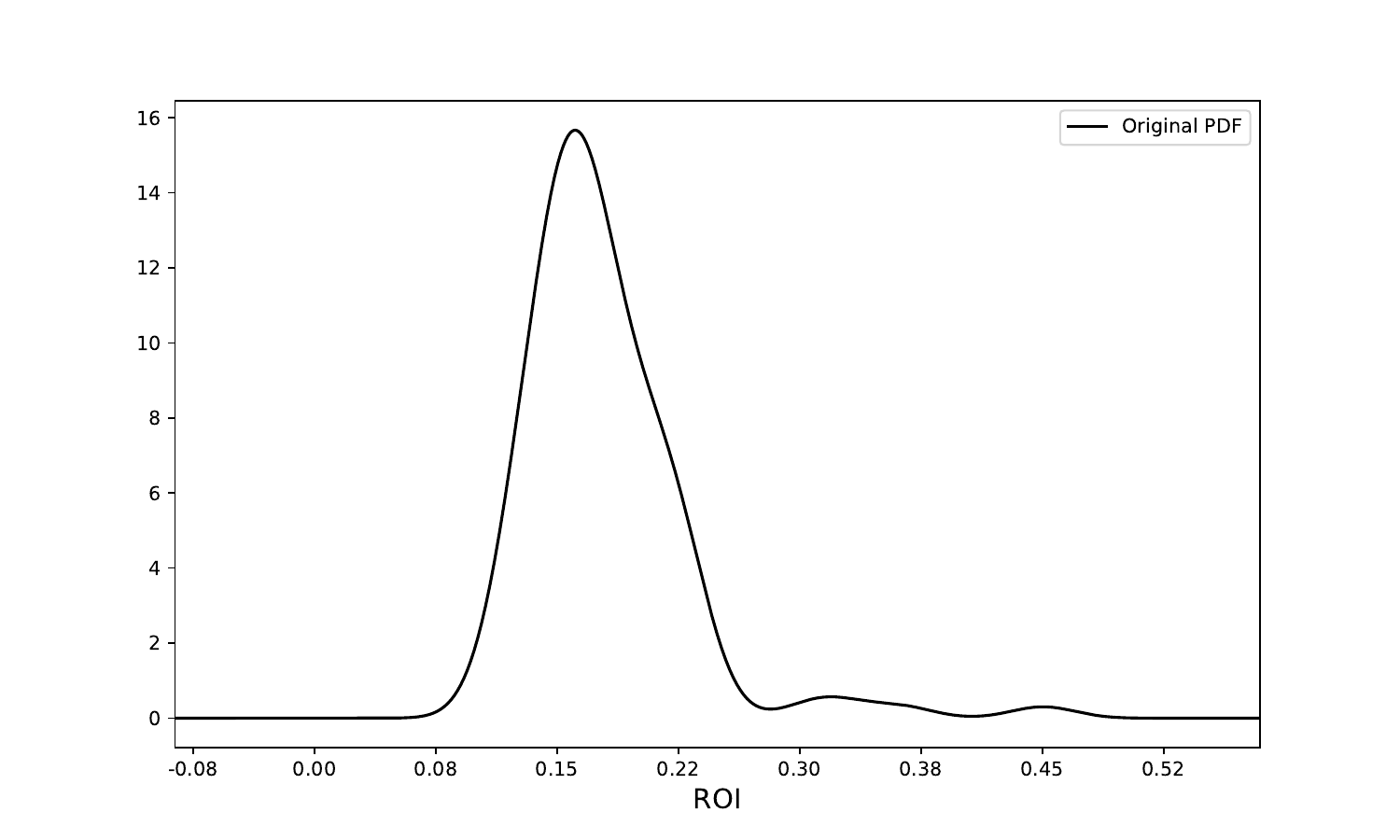}
        \caption{Original gain PDF (after classical PO).}
        \label{fig:losspdforiginal}
    \end{minipage}
\end{figure}

The gain function is defined in terms of ROI and risk is measured using the CVaR-deviation. The statistical data for the assets is synthetic and comprises 100 scenarios, generated by modeling long-term scenarios of assets features (typically a few years). The equality constraint in eq. (\ref{eq:x_eq}) is originally considered, where the budget $B$ is expressed in GigaWatts, and $y_n$ denotes the cost of buying one GigaWatts unit of asset $A_n$.
After a change of variables, the decision variable becomes $\vec P$ and the equality constraint is reformulated as shown in eq. (\ref{eq:p_eq}). 
The constraints in eq. (\ref{eq:constr}) are limited to inequalities per asset, i.e. to a diagonal $\underline M \,\ \underline Y$ matrix, and contain budget information:
\begin{eqnarray}\label{fig:con_res}
    (\underline M \,\ \underline Y)_{n,n} \,\ p_n \leq c_n.
\end{eqnarray}
After classical PO, we obtain portfolio shares $\vec P_a^*$ tagged `original portfolio' in the following, see Fig. \ref{fig:losspdf1}, and the corresponding estimated gain PDF $\hat\sigma(\vec P,.)$ tagged `original PDF' in the following, see Fig. \ref{fig:losspdforiginal} (a standard gaussian kernel is used for $K(.)$ in eq. (\ref{eq:estim})).

\begin{figure}
    \centering
    \includegraphics[width=0.8\linewidth]{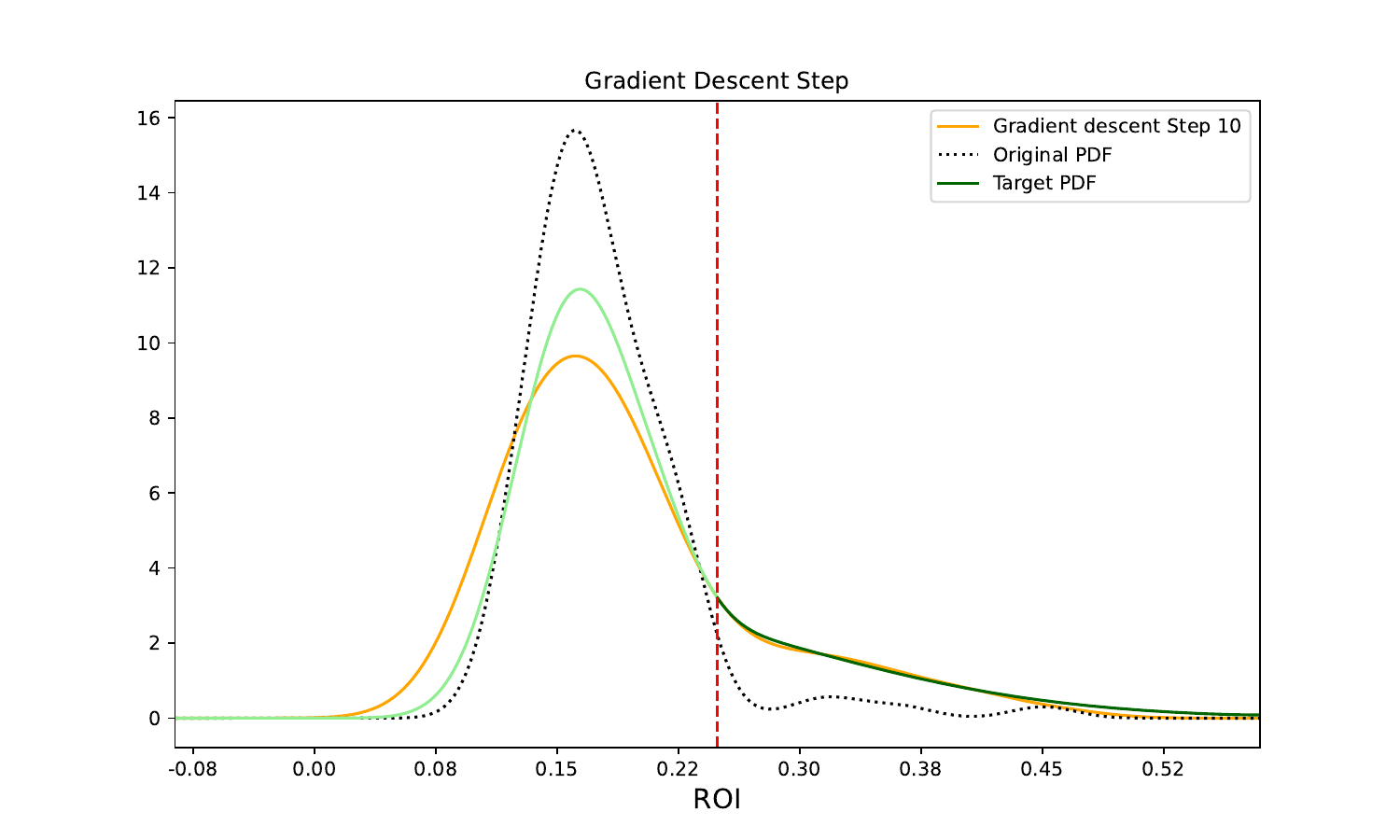}
    \caption{Gain PDF before and after our new optimization, and target PDF. The red dashed line represents the center of a sharp sigmoid. Case where the inequality constraints have been removed.}
    \label{fig:losspdf7}
\end{figure}

Firstly, we show that the method proposed in sections \ref{sec:meth} and \ref{sec:app}
is robust and makes it possible to match very well a target PDF.
To that aim, we relax the inequality constraints in eq. (\ref{fig:con_res}) or in other terms consider $c_n\rightarrow \infty$, and initialize with the original gain PDF the projected gradient iterations that will compute a perturbed PDF.
The target PDF is constructed from the original gain PDF by enhancing the high gains and appears in green in Fig. \ref{fig:losspdf7}. The goal is not to match the entire PDF, but only the part located to the right of the red line (in dark green) in order to favor high gains. 
This is formalized through the use of an L2 norm for \( D(\hat\sigma (\vec P,.); \sigma_t(.)) \) as in eq. (\ref{eq:L2bis}), where the weight \( \theta(.) \) is a steep sigmoid centered on red dashed line in Fig. \ref{fig:losspdf7} so that mostly only the dark green part of the target PDF is encouraged to be matched (we use the sigmoid and not a step function merely to have a differentiable function).
Note that the used target PDF represents a large `perturbation', unrealistic in operational situations but here the goal is to make the test challenging.
The PDF result after projected gradient descent appears in orange in Fig. \ref{fig:losspdf7}, and it near-perfectly matches the dark green part of the target PDF.
This illustrates the robustness of the algorithm but is not fully coherent with the original PO problem as the the per asset inequality constraints were removed.

\begin{figure}
    \centering
    \includegraphics[width=0.8\linewidth]{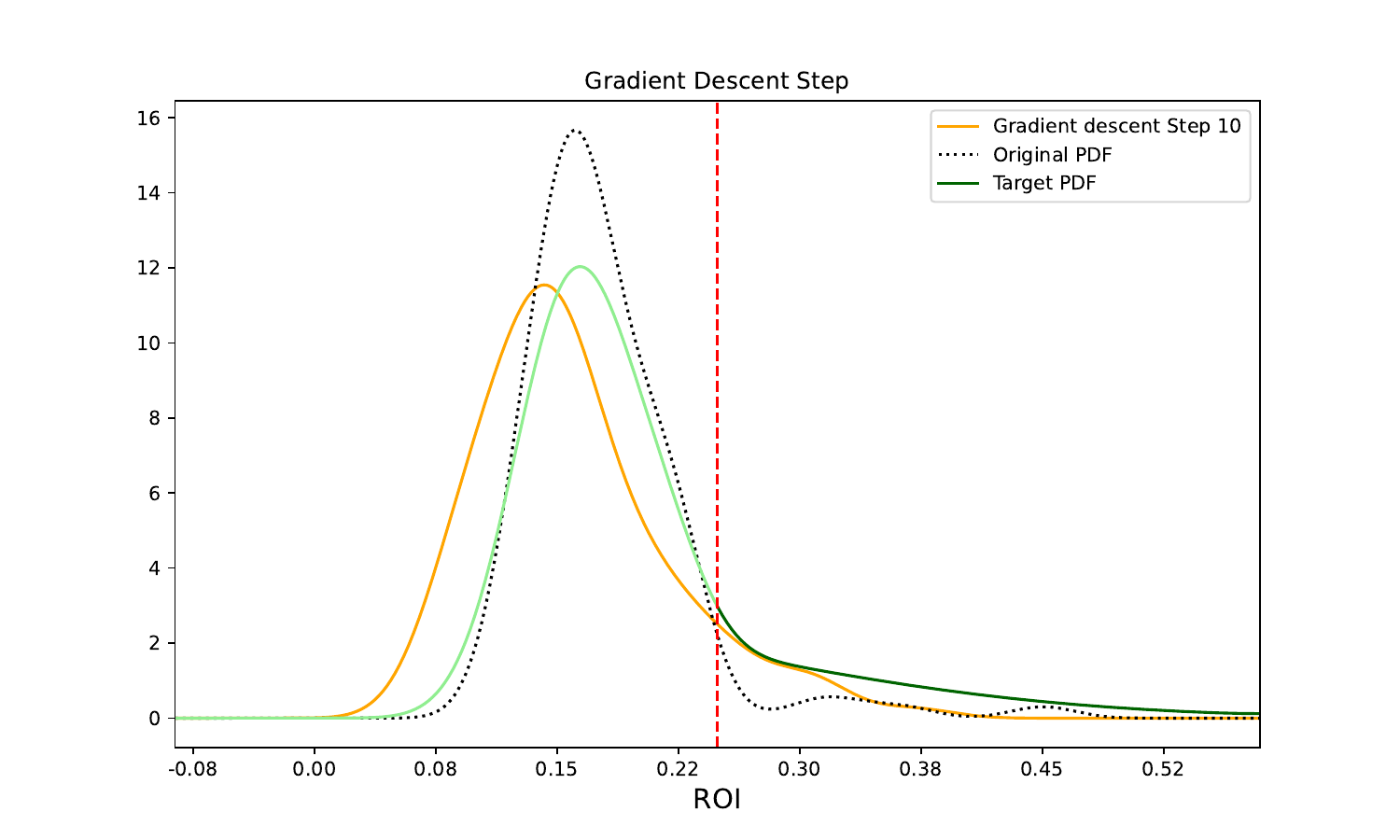}
    \caption{Gain PDF before and after our new optimization, and target PDF. The red dashed line represents the center of a sharp sigmoid. Case with the inequality constraints.}
    \label{fig:losspdf2}
\end{figure}

\begin{figure}
    \centering
    \includegraphics[width=0.95\linewidth]{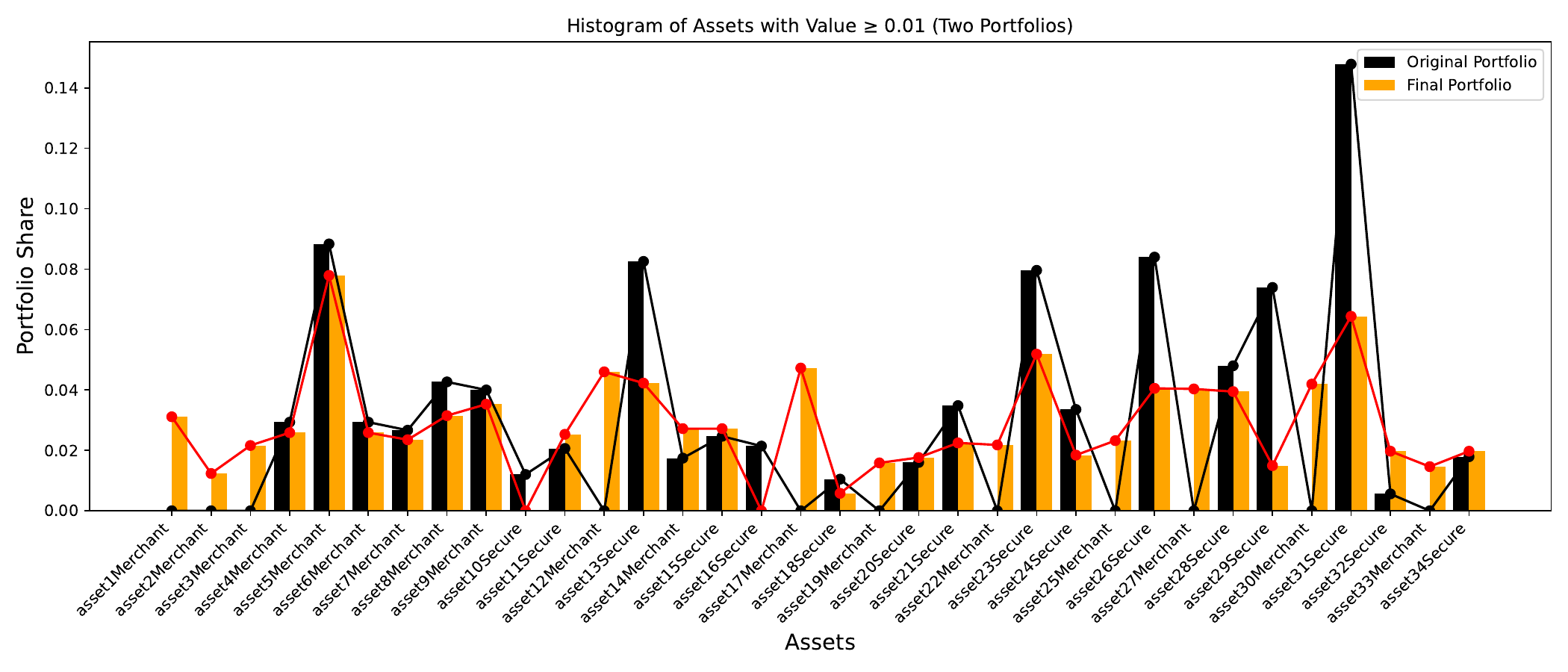}
    \caption{Portfolio shares resulting from conventional portfolio optimization (black) and from the new optimization approach (orange).}
    \label{fig:losspdf3}
\end{figure}

Figure \ref{fig:losspdf2} shows similar results but this time including the inequality constraints. In this case the optimized PDF (in orange) does not exactly match the target PDF  (the dark green part), but approaches it as closely as the constraints allow, demonstrating that the method remains robust even when the target PDF is not precisely defined. This is a key for an application in operation contexts. The corresponding portfolio shares are shown in orange in Fig. \ref{fig:losspdf3} and the initial portfolio shares are shown in black.
It can be observed that
our method fostering high gains suggests increased investment in Merchant assets (with greater variability), thereby enabling the achievement of higher ROI, while investment in Secure assets (with lower variability and thus lower ROI) is reduced, which is fully coherent.



Finally, we apply the marginal cost method described in section \ref{sec:marg_cost} to quantify a budget variation that can be associated to the sub-optimality of the portfolio obtained with our new approach. In this study, the marginal cost variation $\Delta B_t$ was computed by explicitly evaluating the objective function for a range of budget values and identifying the budget increment that yields a similar objective function between the original portfolio and the one obtained with the new method, rather than using a derivative-based approach, since the considered portfolio `perturbation' is not small. We obtain $\Delta B_t=-45$ GigaWatts, meaning that we need to increase the budget of the original PO problem by 45 GigaWatts (and not decrease it). This outcome is not the most common but can occur. Indeed, here more inequality constraints become saturated as we increase the budget $B$ in GigaWatts, forcing the allocation of resources to riskier assets and a decrease of the value of the objective function. The ROI gain function behavior also contributes to this result but the explanation goes beyond the scope of this article (in contrast with a linear gain function case where less restrictive constraints typically result in an increase in the objective function as the budget grows). 

\section{Conclusion}

We have presented a unified framework for PO based on the gain PDF. This approach allows for a more comprehensive characterization of portfolio performance by leveraging the full statistical information available on the gain distribution, rather than relying only on expectation-based objective function terms. The method enables portfolio managers to directly target specific features of the gain distribution, providing maximal flexibility in expressing their preferences and requirements.

We explicated the statistical foundations of the proposed framework, generalized conventional risk measures to accommodate non-linear gain functions, introduced a novel procedure that matches a user-defined target gain PDF and detailed the estimation techniques.

We developed a method to constrain high-gains leveraging the information present in the gain PDF. 
The method proceeds in two stages. First, a conventional PO is performed select a portfolio according to a balance between expected risk and profit. Starting from this portfolio, our new methodology is then applied to further optimize the portfolio by matching a target gain PDF, specifically to foster higher gains. This two-step process allows portfolio managers to benefit from established risk–profit trade-offs while also perturbing the portfolio according to their specific objectives regarding the gain distribution.
A projected gradient algorithm has been proposed for the optimization, to respect operational constraints, as well as a methodology to quantify a marginal cost associated to the perturbed portfolio in a common budget unit, providing a meaningful information to portfolio managers.
The practical relevance of the approach was illustrated through numerical experiments considering an energy-producing assets portfolio.
But the versatility of the proposed approach extends beyond this specific use case (trading, financial assets...). 

Depending on the portfolio manager's objectives, other regions of the gain PDF  can be targeted and optimized, such as loss-side details (for loss control), central region (for stability), or 
multi-modal features. 
Similarly, the marginal cost method extends to any kind of portfolio optimality deviations,
e.g. to evaluate the marginal cost of modifying inequality constraint  parameters, allowing to homogeneously compare heterogeneous variations of parameters and features of the PO problem.
This flexibility makes the methodology suitable for a wide range of applications and contexts. Future work will among others explore automated target PDF construction and develop numerical experiences on various PO applications and datasets.






\section*{Acknowledgment}

We would like to thank Matthieu Blondel, Thomas Balsan, Emeric de Monteville, David Jambois and Mikael Soliveres for their feedback and insightful comments on the methodology described in this article. 
We thank TotalEnergies for the permission to publish this work.








\bibliography{sn-bibliography}

\end{document}